\documentstyle[12pt,epsf,epsfig]{article}
\def\be{\begin{equation}}
\def\ee{\end{equation}}
\def\ba{\begin{eqnarray}}
\def\ea{\end{eqnarray}}
\def\la{~\mbox{\raisebox{-.6ex}{$\stackrel{<}{\sim}$}}~}
\def\ga{~\mbox{\raisebox{-.6ex}{$\stackrel{>}{\sim}$}}~}
\def\bq{\begin{quote}}
\def\eq{\end{quote}}

 at 10truept

\newcommand{\beq}{\begin{equation}}
\newcommand{\eeq}{\end{equation}}
\newcommand{\beqa}{\begin{eqnarray}}
\newcommand{\eeqa}{\end{eqnarray}}

\def\la{~\mbox{\raisebox{-.6ex}{$\stackrel{<}{\sim}$}}~}
\def\ga{~\mbox{\raisebox{-.6ex}{$\stackrel{>}{\sim}$}}~}

\def\ltap{\ \raise.3ex\hbox{$<$\kern-.75em\lower1ex\hbox{$\sim$}}\ }
\def\gtap{\ \raise.3ex\hbox{$>$\kern-.75em\lower1ex\hbox{$\sim$}}\ }
\def\gl{\ \raise.5ex\hbox{$>$}\kern-.8em\lower.5ex\hbox{$<$}\ }
\def\roughly#1{\raise.3ex\hbox{$#1$\kern-.75em\lower1ex\hbox{$\sim$}}}

\begin{document}
\thispagestyle{empty}
\begin{flushright}
hep-ph/0312002\\ November 2003
\end{flushright}
\vspace*{1cm}
\begin{center}
{\Large \bf Disformal Inflation}\\
\vspace*{1.5cm}
{\large Nemanja Kaloper\footnote{\tt kaloper@physics.ucdavis.edu} }\\
\vspace{.5cm}
{\em Department of Physics, University of California, Davis, CA 95616}\\
\vspace{.15cm}
\vspace{2cm} ABSTRACT
\end{center}
We show how short inflation naturally arises in a non-minimal
gravity theory with a scalar field without any potential terms.
This field drives inflation solely by its derivatives, which
couple to the matter only through the combination $\bar g_{\mu\nu}
= g_{\mu\nu} - \frac{1}{m^4} \partial_\mu \phi \partial_\nu \phi$.
The theory is free of instabilities around the usual Minkowski
vacuum. Inflation lasts as long as $\dot \phi^2 > m^4$, and
terminates gracefully once the scalar field kinetic energy drops
below $m^4$. The total number of e-folds is given by the initial
inflaton energy $\dot \phi_0^2$ as ${\cal N} \simeq \frac13
\ln(\frac{\dot \phi_0}{m^2})$. The field $\phi$ can neither
efficiently reheat the universe nor produce the primordial density
fluctuations. However this could be remedied by invoking the
curvaton mechanism. If inflation starts when $\dot \phi^2_0 \sim
M^4_P$, and $m \sim m_{EW} \sim TeV$, the number of e-folds is
${\cal N} \sim 25$. Because the scale of inflation is low, this is
sufficient to solve the horizon problem if the reheating
temperature is $T_{RH} \ga MeV$. In this instance, the leading
order coupling of $\phi$ to matter via a dimension-8 operator
$\frac{1}{m^4}\partial_\mu \phi
\partial_\nu \phi ~ T^{\mu\nu}$ would lead to
fermion-antifermion annihilation channels $f\bar f \rightarrow
\phi \phi$ accessible to the LHC, while yielding very weak
corrections to the Newtonian potential and to supernova cooling
rates, that are completely within experimental limits.

\vfill
\setcounter{page}{0}
\setcounter{footnote}{0}
\newpage

The recurring challenge to our attempts to understand Nature is
the origin of hierarchies between the scales we observe. Familiar
examples are the hierarchy between the Planck scale $M_P \sim
10^{19} GeV$ and the electroweak scale $m_{EW} \sim TeV$,
$M_P/m_{EW} \sim 10^{16}$, and the hierarchy between the Planck
scale and the present horizon scale, $H_0 \sim 10^{-33} eV$,
$M_P/H_0 \sim 10^{61}$. These problems are usually dealt with
separately. In the former case, models of particle dynamics such
as strong gauge field dynamics \cite{tech,lhiggs}, supersymmetry
\cite{susy} or large extra dimensions \cite{led,rs} are invoked to
explain the dichotomy between the Planck and electroweak scales.
In the latter case, the leading contender to explain the horizon
scale is inflation \cite{inflation}, which posits that the
universe has been blown up really large by a period of exponential
expansion in the past, and then subsequent expansion generates the
rest of the hierarchy between $M_P$ and $H_0$. If inflation starts
near the Planck scale, it should blow up the universe by at least
${\cal N}_* \sim 65$ e-folds, or by a factor of at least $e^{{\cal
N}_*} \sim 10^{28}$. The approximate relation $e^{{\cal N}_*} \sim
(M_P/m_{EW})^2$ is typically viewed as an accident. In fact, the
usual models of inflation predict that the universe has expanded
by much more than the current necessary minimum to explain the
present horizon scale \cite{chaotic}. This would indicate that
there is nothing special about the present horizon scale. We just
happen to make our observations now, but some other being could
have seen a completely different horizon scale at some other time,
as the cosmic evolution marches on.

Yet there are indications that we might live at a special moment
in the history of the universe. Indeed observations have uncovered
the cosmic coincidences: the current cosmological densities of
various forms of matter inhabiting our universe, such as dark
energy, dark matter, baryons, photons and neutrini are within a
few orders of magnitude of each other \cite{coincidences}. Some of
the coincidences are presently very mysterious, such as explaining
the scale of dark energy from first principles. Other
coincidences, such as the near equality of the energy densities of
dark matter, baryons and photons, may be understood in particle
physics models which contain weakly-interacting particles with
masses and couplings set by the electroweak scale $m_{EW}$. Any
definitive clue in favor of spatial curvature within a few orders
of magnitude of the critical density of the universe would further
underscore that we live in a special epoch, requiring that
inflation were short. It should have ceased after the necessary
minimum of e-folds was achieved, in order to avoid completely
flattening the spatial slices. Other clues of short inflation
might emerge from observing non-trivial topology of the universe
\cite{topology}, low power in low $\ell$ CMB multipoles
\cite{power}, substructure in the CMB \cite{cmb} or holographic
considerations \cite{holo}. A natural explanation for such
coincidences would be to relate the dynamics which control their
evolution, including cosmology, with a particular hierarchy of
scales governing microphysics, such as $M_P/m_{EW}$.

Building models of inflation capable of stopping after few tens of
e-folds has been especially hard (for some models see
\cite{open,tom,andrei,giashamit}). In this note, we consider a
mechanism where inflation can be very short. The inflaton is a
massless singlet pseudoscalar, whose dynamics respects the shift
symmetry $\phi \rightarrow \phi + {\cal C}$ and reflection, $\phi
\leftrightarrow - \phi$. Its couplings to the matter sector are
introduced via a modification of the gravitational coupling to
matter, of the form
\be
\bar g_{\mu\nu} = g_{\mu\nu} - \frac{1}{m^4}
\partial_\mu \phi \partial_\nu \phi \, .
\label{metrics}
\ee
Here the metric $g_{\mu\nu}$ is the canonically normalized metric
with the kinetic term given by the usual Einstein-Hilbert action,
and $\phi$ is normalized as usual such that it has dimension of
mass. The mass scale $m$ is the coupling parameter of the inflaton
sector to matter, which couples covariantly to the combination
$\bar g_{\mu\nu}$. We will discuss the acceptable range of values
for it below. Theories with scalars coupled to matter in ways
including (\ref{metrics}) have been considered by Bekenstein in
1992 \cite{bekenstein}, who looked for generalizations of
Riemannian geometry that do not violate the weak equivalence
principle and causality. He found that the extensions of the
standard General Relativity based on coupling the matter to the
combinations of the form he referred to as {\it disformal}
transformation,
\be \bar g_{\mu\nu} ={\cal A}(\phi, (\partial \phi)^2) g_{\mu\nu}
- \frac{{\cal B}(\phi, (\partial \phi)^2)}{m^4}\partial_\mu \phi
\partial_\nu \phi \,
\label{bekmetrics}
\ee
preserve causality and the weak equivalence principle. In contrast
to conformal transformation, the disformal transformation
(\ref{bekmetrics}) does not preserve the angles between the
geodesics of $g_{\mu\nu}$ and $\bar g_{\mu\nu}$. We confine our
attention to a specialized form (\ref{metrics}), taking ${\cal A}=
{\cal B} =1$ (constants other than unity can be absorbed away by
rescaling $M_P$ and $m$), in order to enforce the symmetries $\phi
\rightarrow \phi + {\cal C}$, $\phi \leftrightarrow - \phi$ which
protect the inflaton from the matter loop corrections. This
implies the stability of the slow roll regime under the Standard
Model radiative corrections.

With the choice of the mass scale $m \sim m_{EW} \sim TeV$, the
resulting dynamics is equivalent to low scale inflation with
$V^{1/4} \sim TeV$, lasting about $25$ e-folds \cite{lowscale}.
This is just enough to solve the horizon problem if the reheating
is $T_{RH} \ga MeV$ \cite{kolbturner}. The reheating and the
generation of density perturbations are however involved. The
field $\phi$ which drives inflation cannot efficiently reheat the
universe, nor produce the scale-invariant spectrum of
perturbations to match the COBE amplitude, because it is too
weakly coupled to the Standard Model, and the scale of inflation
is so low. The model also does not solve the curvature problem,
because it requires the initial curvature of the universe to be
small in order not to prevent the onset of the low scale
inflation. However these problems are common in low scale
inflation. The reheating and the generation of density
perturbations may be solved by invoking a curvaton field
\cite{curv}. We will outline a scenario that could accomplish
this. Solving the curvature problem requires additional dynamics,
such as a stage of very early inflation \cite{ahdkmr} or
holographic considerations \cite{tomwilly}.

We define the theory by the action principle $\delta S = 0$, where
the action is
\be
S=\int d^4x \Bigl\{\sqrt{g}\bigl[\frac{M^2_P}{2}
R - \frac12{(\partial \phi)^2} \bigr] - \sqrt{\bar g}{\cal
L}_M(\psi,
\partial \psi,\bar g^{\mu\nu}) \Bigr\} \, ,
\label{action}
\ee
where $g = \det(-g_{\mu\nu})$ etc. Because of the shift symmetry
of $\phi$, general covariance and reflection $\phi \leftrightarrow
- \phi$ we can treat the matter Lagrangian ${\cal L}_M$ as fully
quantum, including all the Standard Model loop corrections. The
shift symmetry operates as in the case of pseudo-Nambu-Goldstone
inflatons \cite{natural,pseudon}, excluding corrections which are
polynomial in $\phi$. The reflection $\phi \leftrightarrow - \phi$
precludes the operators of the form $\partial_\mu \phi j^\mu$
where the scalar couples derivatively to some conserved current.
Finally, general covariance of the matter sector protects the
universality of matter couplings to only $\bar g^{\mu\nu}$, which
can be seen by rewriting the action (\ref{action}) in terms of
only barred variables and recalling that by matter loops we mean
these loop diagrams which involve only matter internal lines.
Specifically, the Standard Model corrections do not change the
coupling constant ${1}/{m^4}$. Varying (\ref{action}) yields the
field equations, which using the shorthand $U_\mu =
{\frac{1}{m^2}}
\partial_\mu \phi$ are
\ba
&& M^2_P G^{\mu\nu} = \partial^\mu \phi \partial^\nu \phi -
\frac12 g^{\mu\nu} (\partial \phi)^2
+ \sqrt{1-U^2} ~{\bar T}^{\mu\nu} \, ,  \label{fieldeq1} \\
&&{\bar \nabla}_\mu {\bar T}^{\mu\nu} = 0 \, ,  \label{fieldeq2} \\
&& \nabla^2 \phi + \frac{1}{m^4} \sqrt{1-U^2} ~{\bar T}^{\mu\nu}
{\bar \nabla}_\mu {\bar \nabla}_\nu \phi = 0 \, . \label{fieldeq3}
\ea
The equation (\ref{fieldeq1}) is the modified Einstein's equation,
(\ref{fieldeq2}) stands for the matter field equations,
designating that the matter fields couple to $\bar g_{\mu\nu}$,
and (\ref{fieldeq3}) the inflaton field equation, which includes
the matter-inflaton derivative couplings. Raising and lowering of
the indices of unbarred tensors is to be done with
$\Bigl(g^{\mu\nu}, g_{\mu\nu}\Bigr)$, and of barred tensors with
$\Bigl( \bar g^{\mu\nu}, \bar g_{\mu\nu} \Bigr)$. It is
straightforward to derive several useful relations between key
barred and unbarred quantities; using (\ref{metrics}), one finds
$\bar g =(1 - { U}^2) g$, $\bar g^{\mu\nu} = { g}^{\mu\nu} +
\frac{1}{1 - { U}^2} { U}^\mu { U}^\nu$, ${\bar U}^2 =\frac{{
U}^2}{1-{ U}^2}$, ${\bar U}^\mu = \frac{1}{1-{U}^2} {U}^\mu$.

Before proceeding we ought to mention that the theories of this
form have been considered in the context of the so called variable
speed of light cosmologies \cite{vsl}. The motivation was to argue
that if $\partial \phi \ne 0$, the lightcones of the metrics
$g_{\mu\nu}$ and $\bar g_{\mu\nu}$ are different, suggesting that
the electromagnetic waves propagate faster than gravity waves,
with a speed which varies in space and time. This ``superluminal"
propagation of light is then supposed to solve the horizon problem
without inflation, since it would seem to allow for communication
at superhorizon scales. We strongly caution against considering
the theory (\ref{metrics}), (\ref{action}) in this way. Namely,
because the $\phi$-field equation (\ref{fieldeq3}) is homogeneous
in $\partial \phi$, it admits solutions $\phi = {\rm const}$,
which are identical to $\phi = 0$ by the shift symmetry. This is
the vacuum of the theory. In this vacuum there is no difference in
the propagation speed of any excitations in the theory, matter or
gravitational. Thus the presence of two different lightcone
structures, one for the graviton and another for the matter
fields, is an environmental effect, which emerges because the
initial state of the universe began with $\partial \phi \ne 0$.
This is analogous to the propagation of light in a dielectric, or
to the propagation of massless charged particles in an external
electric field. An observer who sees that the trajectories of
these probes deviate from the null geodesics in the vacuum does
not invoke a changing speed of light at a fundamental level to
explain this. Instead she notes that the probes interact with the
environment, which breaks Poincare symmetry because $\partial \phi
\ne 0$. The breaking is soft, in the sense that as $\partial \phi$
diminishes in the course of the evolution of the universe, the
symmetries are restored. This is reminiscent to a spontaneously
broken gauge symmetry, where because of the breaking the gauge
fields become massive, and their quanta propagate along timelike
instead of null geodesics. Because in this case the scalar field
gradients $\partial \phi \ne 0$ break Poincare symmetry instead of
the electromagnetic gauge symmetry, the photons remain massless
and move along null geodesics, while the gravitons move along
timelike geodesics. One ought to interpret the double lightcone
structure induced by $\partial \phi \ne 0$ as a signature of the
slow-down of gravitons due to their strong interactions with
$\partial \phi$, which makes the early universe opaque to them.
This helps with the horizon problem not because it allows for
superhorizon correlations, but because it arrests the
gravitational instability, preventing the growth of
inhomogeneities. However in the frame where the matter fields are
canonically normalized this looks precisely like inflation. Hence
in what follows we adopt this view and focus on the effective
field theory description of inflation.

Let us now establish when the model based on
(\ref{action})-(\ref{fieldeq3}) is meaningful. Consider first the
low energy limit. As indicated above, we define the vacuum by
setting $\phi=0$. For simplicity we further assume that in the
vacuum $\bar T_{\mu\nu} = 0$ and so $g_{\mu\nu}=\eta_{\mu\nu}$,
i.e. that the vacuum is the usual Minkowski space. In order to
ensure its perturbative stability we must show that it is a
minimum energy state, without negative energy excitations and/or
runaway modes (i.e. ghosts and tachyons). Constructing the matter
sector in the usual way ensures that there are no such degrees of
freedom in ${\cal L}_M$. The form of the gravitational action in
(\ref{action}) further guarantees that the metric degrees of
freedom are safe too. What remains to check is that the scalar
$\phi$ does not produce instabilities. Now, if we consider small
perturbations of (\ref{fieldeq3}) around the vacuum $\phi = \bar
T_{\mu\nu} = 0$, $g_{\mu\nu} = \eta_{\mu\nu}$, we see that the
scalar $\phi$ is just a massless canonically normalized scalar
field too, without any pathologies. Thus the vacuum is stable.

However there still might be runaway scalar modes around some
fixed classical background with $\bar T_{\mu\nu} \ne 0$. Even if
the vacuum were exactly stable, it would be disastrous if
infinitesimally small distributions of matter are not. To check
this does not occur we consider the spectral decomposition of
$\phi$ in the presence of a point mass. This will be sufficient
since any other distribution of energy-momentum can be obtained by
superposition and boosting of such sources. Before looking at the
details, however, we note that the dimensionless coefficient
controlling the correction is given by $\sim \rho/m^4$, where
$\rho$ is the energy density of the distribution. If we smooth the
distribution over a whole Hubble volume, this reaches its upper
value if the total mass $M$ is of the order of the mass in the
observable universe: $M \sim \rho_0/H^3_0 \sim M^2_P/H_0$, which
is at most $\sim \frac{M^2_P H^2_0}{m^4}$. This is smaller than
unity as long as $m > 10^{-3} eV$. In fact we will see below that
$m$ is at least $m_{EW}$, and so the perturbation is really tiny,
with $\xi$ at most $10^{-60}$. The parameter $\xi$ approaches
unity only in the limit $\rho \rightarrow m^4$. From this we
expect that the $\phi$ excitations will not destabilize the
background as long as the densities are below $m^4$. Similar
conclusions remain true for localized sources too. To see this
explicitly, we expand (\ref{fieldeq3}) around a point mass.
Picking $\phi = 0$ and $g_{\mu\nu} = \eta_{\mu\nu}$ for the
background outside of the mass source, we find the equation for
the excitations of $\phi$,
\be
\partial^2 \phi + \frac{1}{m^4} M \delta^{(3)}(\vec x)
\ddot \phi = 0 \, , \label{pertfieldeq3}
\ee
where $M$ is the mass of the source at $\vec x = 0$ and
$\delta^{(3)}(\vec x)$ is the Dirac $\delta$-function. Using
(\ref{pertfieldeq3}), after simple algebra we can write the matrix
propagator equation for $\Delta(\omega, \vec k) = i \langle
\phi(\omega, \vec k) \, \phi(0) \rangle$ in momentum space,
\be \Bigl(\vec k^2 - \omega^2 (1 - \xi) \Bigr) \Delta(\omega,\vec
k) + \xi \omega^2 \sum_{\vec q \ne \vec k} \Delta(\omega, \vec q)
= - i \, , \label{pertprop} \ee
where $\xi = \frac{M H_0^3}{m^4} = \frac{M H_0}{M^2_P} \frac{M^2_P
H^2_0}{m^4} \ll 1$, and we imagine that the universe is a lattice
of size $1/H_0$ with a lattice spacing $1/\Lambda$. We can solve
the equation (\ref{pertprop}) perturbatively using $\xi$ as the
expansion parameter, to find
\be \Delta(\omega, \vec k) = \frac{i}{\omega^2(1-\xi) - \vec k^2 +
i \epsilon } \Bigl( 1 + \xi \sum_{\vec q\ne \vec k}
\frac{\omega^2}{ \omega^2(1-\xi) - \vec q^2 + i \epsilon} + \ldots
\Bigr) \, . \label{solpertprop} \ee
This shows that the full propagator in the presence of a mass
source $M$ contains admixtures of all plane wave modes with very
slightly shifted frequencies $\omega^2 \rightarrow \omega^2
(1-\xi)$. However when $\xi \ll 1$ all the poles occur only when
$\omega^2 > 0$, and thus there are no runaway, exponentially
growing modes. Moreover the momenta on the lattice are $\vec p =
H_0 \vec n$, where $\vec n \in {\tt Z}^3$, and therefore at the
poles $\omega^2(1-\xi) - \vec q^2 = \vec p^2 - \vec q^2 = H^2_0
(\vec n_p^2 - \vec n_q^2)$. Hence $\xi \frac{\omega^2}{
\omega^2(1-\xi) - \vec q^2 + i \epsilon}$ is maximized when $\vec
p^2 - \vec q^2 \simeq |\vec n| H^2_0 \simeq H_0 \omega$, reaching
$\xi \omega/H_0 \le \xi \Lambda/H_0$. Hence as long as the theory
is cut off at a scale $\Lambda \le H_0/\xi \la M_P$ the residues
are positive, and so there are no negative energy excitations
either. Thus the Minkowski vacuum of the theory
(\ref{action})-(\ref{fieldeq3}) is perturbatively stable.

These conclusions are valid as long as the energy density of the
Standard Model matter in $\bar T_{\mu\nu}$ does not exceed ${\cal
O}(m^4)$. As it increases towards $m^4$, the expansion parameter
$\xi$ approaches unity and the perturbative analysis yielding
(\ref{solpertprop}) breaks down. This is not necessarily
detrimental: it means that the theory based on (\ref{action}) must
be given a proper UV completion. Thus to ensure the validity of
the effective field theory description of $\phi$ as defined by
(\ref{action}) we should cut off the matter sector at
$\lambda_{SM} \sim m$. Once this is done, the $g_{\mu\nu}, \phi$
sector may remain well-defined all the way up to some high energy
scale $\Lambda \sim M_P$ which regulates the gravity-$\phi$
sector. Such frameworks were discussed in, for example,
\cite{adg}, who suggested that the Standard Model is completed by
a $TeV$-scale Little String Theory, which couples to gravity that
remains weak up to the usual Planck scale. Note that although we
imagine that $\partial \phi$ can reach energy scales as high as
$M^2_P$, this does not destabilize the Standard Model sector
because it couples to $\phi$ only through $\bar g^{\mu\nu} = {
g}^{\mu\nu} + \frac{1}{1 - (\partial \phi)^2/m^4} {\partial}^\mu
\phi {\partial }^\nu \phi /m^4$. Therefore the dependence on the
cutoff $\Lambda$ cancels to the leading order, entering only
through terms $\sim m^4/\Lambda^4$, leaving $m$ in full control of
the Standard Model as long as we ignore gravity and $\phi$ loops.

The inclusion of the Standard Model corrections to the
$g_{\mu\nu}, \phi$ sector does not destabilize the leading order
terms in $g_{\mu\nu}, \phi$ in (\ref{action}). Because the
Standard Model is cut off at $m$, and because it only couples to
$\bar g_{\mu\nu}$, general covariance implies that the Standard
Model corrections are organized as an expansion in the
higher-derivative invariants of $\bar g_{\mu\nu}$. The only
dimensional scale weighing them is $m$:
\be {\cal L}_{corrections} = \sqrt{\bar g} \Bigl( a_0 m^4 + a_1
m^2 \bar R + a_2 \bar R^2 + a_3 {\bar \nabla}^2 \bar R +
\frac{a_4}{m^2} \bar R^3 + \ldots \Bigr) \, , \label{corrections}
\ee
where the coefficients $a_0, a_1, a_2, \ldots$ are all numbers of
order unity. We can now add the leading order terms for
$g_{\mu\nu}$ and $\phi$ from (\ref{action}),
$\sqrt{g}\bigl[\frac{M^2_P}{2} R - \frac12{(\partial \phi)^2}
\bigr]$. The full effective Lagrangian rewritten in terms of the
variables $g_{\mu\nu}$ and $\phi$ becomes symbolically
\ba {\cal L}_{eff} &=& \sqrt{\bar g} \Bigl( \frac{1}{\sqrt{1-
(\bar
\partial \phi)^2/{m^4}}} \bigl[\frac{M^2_P}{2} \bar R - \frac{M^2_P
(\bar
\partial \phi)^2}{m^4} \bar R - \frac12{(\bar
\partial \phi)^2} \bigr] \nonumber \\
&& + a_0 m^4 + a_1 m^2 \bar R + a_2 \bar R^2 + a_3 {\bar \nabla}^2
\bar R + \frac{a_4}{m^2} \bar R^3 + \ldots \Bigr) \, ,
\label{corrs} \ea
where we have ignored the tensor structure in the terms like
$\frac{M^2_P}{m^4} \partial_\mu \phi \partial \phi_\nu \bar
R^{\mu\nu}$, choosing to write them instead as $\frac{M^2_P}{m^4}
(\bar \partial \phi)^2 \bar R$, which is sufficient to analyze
their scaling, and relative importance in the effective action
with the Standard Model corrections included. When $\partial \phi<
m^2$, the corrections are obviously small. In the regime $\partial
\phi \sim \Lambda^2$, in the background (\ref{solns}) each
derivative contributes a power of $\bar H \simeq m^2/M_P$, and so
the expansion becomes a series of the form
\be {\cal L}_{eff} = \sqrt{\bar g} \Bigl( m^2 \Lambda^2 + m^4 +
a_0 m^4 + a_1 \frac{m^6}{M^2_P} + a_2 \frac{m^8}{M^4_P} + a_3
\frac{m^8}{M^4_P} + a_4 \frac{m^{10}}{M^6_P} + \ldots \Bigr) \, ,
\label{exppowers} \ee
where the leading order terms $\sim m^2 \Lambda^2$ and $\sim m^4$
come entirely from the classical background, and the corrections
affect the background only slightly through the cosmological term
$\sim a_0 m^4$, while all other effects from terms $\propto a_k$
remain completely negligible. We stress however that in general
the corrections from the $g_{\mu\nu}, \phi$ loops are not under
control, and to understand what happens with them one must seek an
embedding of the theory (\ref{action}) into some more fundamental
theory with a UV completion which is under control. That task is
beyond the scope of the present work. We do see however, that like
in natural inflation scenarios \cite{natural}, that the conditions
for slow roll regime are protected from the matter radiative
corrections.

The presence of a new degree of freedom $\phi$ leads to many new
processes, some of which could affect the low energy experiments.
This yields important observational bounds on $m$. The strongest
arises from collider data. The operator (\ref{vertex}) opens up
the channel for annihilation of any two standard model fermions
into two $\phi$'s, $f\bar f \rightarrow \phi \phi$. The
cross-section for this process goes as
\be \sigma_{f\bar f \rightarrow \phi \phi} \sim \frac{s^3}{m^8} \,
, \label{ancross} \ee
where $\sqrt{s}$ is the center-of-mass energy. Taking $\sqrt{s}
\sim 100 GeV$ and requiring that $\sigma \le \frac{1}{m_{EW}^2}$
in order for this channel not to be ruled out by present data, we
find a bound
\be m \ga m_{EW} \, . \label{gravabound} \ee
If this bound is saturated, the detection of $\phi$'s may be
within reach of the future colliders such as the LHC.

The massless scalar $\phi$ mediates a new force, that could be
long-range, modifying the Newton's law. Even though
(\ref{metrics}) preserves weak equivalence principle, the force
generated by $\phi$ should be constrained by Solar system tests of
gravity just like in the usual Brans-Dicke theory. However in this
case the corrections to the Newton's law are very small, and the
Solar system tests are easy to pass. This can be seen as follows.
We can compute the potential from $\phi$ exchange using Feynman
diagrams. Expanding (\ref{action}) around the vacuum, we find that
the $\phi$-matter interaction vertex is given by the dimension-8
operator
\be {\cal L}_I = \frac{1}{m^4} \partial_\mu \phi \partial_\nu \phi
~ T^{\mu\nu} \, , \label{vertex} \ee
where we can drop the bar from $T^{\mu\nu}$ whenever we expand
around the vacuum. The leading-order diagrams correcting the
Newtonian potential are given in Fig. 1).
\begin{figure}[thb!] \hspace{1.2truecm} \vspace{-1truecm}
\epsfysize=2.8truein \epsfbox{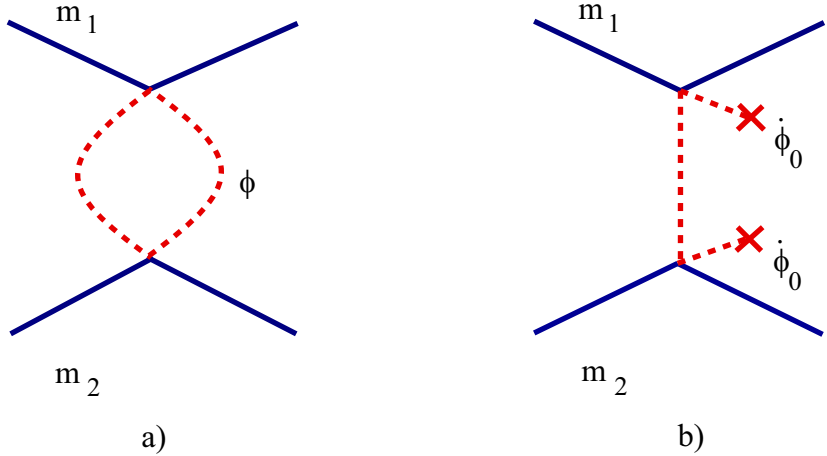}
\caption[]%
{\small\sl Feynman diagrams for the force mediated by $\phi$. }
\end{figure}

The diagram in Fig. 1 a) involves a double $\phi$ exchange. The
loop integral is divergent and so we need to cut it off at some
scale $\Lambda$. The result is the expansion
\be \Lambda^4 + \Lambda^2 \vec k^2 + \vec k^4 \log \vec k^2 +
\ldots \label{expansion} \ee
Both of the cutoff-dependent terms are contact interactions,
corresponding to shrinking both, or one of the propagators in the
loop to a point, and they should be subtracted away, leaving the
$\propto \vec k^4 \log \vec k^2$ term as the physical loop
contribution. This yields
\be V_1 \sim \frac{1}{m^8}  \frac{m_1 m_2 }{r^7} = \frac{1}{M^2_P}
\frac{m_1 m_2 }{r} \frac{M^2_P }{m^8 r^6}\, ,\ee
where the latter parameterization makes the comparison with the
experimental data more transparent. Because of the rapid drop of
this potential with distance, the strongest bounds will come from
the shortest scales that have been probed so far, i.e. from
table-top experiments \cite{tabletop}. Thus taking $r \sim 0.1
{\rm mm}$, we must choose $m$ such that $\frac{M^2_P }{m^8 (0.1
{\rm mm})^6} < 1/100$. We find
\be m^8 \ga 10^8 \, \frac{M^2_P}{{\rm mm}^6} \, ,
\label{gravbound} \ee
or numerically $m>MeV$. Hence as long as $m > MeV$, the force
which $\phi$ mediates is very weak, and short-ranged. In fact, if
we take $m \sim m_{EW}$, which as we will see below is the
strongest bound on $m$, the force becomes strong only at distances
$r \le 40 ~{\rm fermi}$, where the effect would, remarkably,
appear as a sudden opening of six new dimensions. This is very
similar to the theories with large extra dimensions \cite{led} or
CFT effects \cite{ami} in cutoff AdS braneworlds \cite{rs2}.

The diagram in Fig. 1 b) involves a single $\phi$ exchange between
two masses $m_1, m_2$, with two lines ending on the cosmological
background $\dot \phi_0$. The potential arising from this diagram
is, after cancelling the contact terms, the velocity-dependent
contribution to the potential, which arises because the coupling
$\frac{1}{m^4} ~ \partial_\mu \phi \partial_\nu \phi \, T^{\mu\nu}
$ vanishes in the static limit when the mass sources are at rest:
\be V_2 \sim \frac{\dot \phi_0^2}{m^8} \frac{m_1 m_2 }{r^3} \vec
v_1 \cdot \vec v_2  \, .\ee
Because today $\dot \phi_0$ is at most of the order of
$\sqrt{\rho_0} \sim M_P H_0$, $\frac{\dot \phi_0^2}{m^8} <
\frac{M^2_P H^2_0}{m^8} \ll \frac{1}{m^2 M^2_P}$. The bound
(\ref{gravabound}) renders the effects of this term ignorably tiny
at distances $r > m^{-1}$.

The bounds which one obtains from astrophysics considerations are
also consistent with (\ref{gravabound}). They arise because the
coupling of $\phi$ to the matter degrees of freedom via the
dimension-8 operator (\ref{vertex}) leads to the $\phi$ production
which could enhance the cooling rates of astrophysical objects.
The analysis is similar to the one performed in theories with
large extra dimensions \cite{led}. The leading order process that
governs the $\phi$ production is ${\bf p} \rightarrow {\bf p} +
\phi + \phi$, given by the diagram in Fig. 2).
\begin{figure}[thb!]
\hspace{+2truecm} \vspace{-2truecm} \epsfysize=2.4truein
\epsfbox{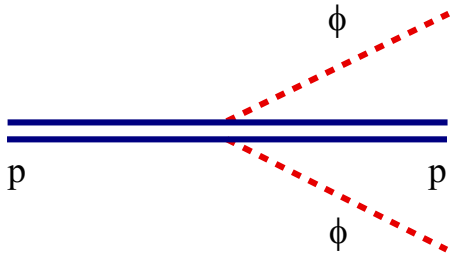}
\caption[]%
{\small\sl The process for the ${\bf p} \rightarrow {\bf p} + \phi
+ \phi$ ``decay". }
\end{figure}
Here ${\bf p}$ is a typical particle in the supernova which is
dressed by thermal effects, as denoted by the double-line in the
Fig. 2). Since it has thermal width it can shake off two $\phi$'s,
decreasing its thermal energy. The ``decay rate" governing this
process is easy to estimate from (\ref{vertex}) simply by
dimensional analysis. Since the typical kinetic energy of a
particle in a star is $E \sim T$, where $T$ is the temperature,
and since the decay rate is proportional to the square of the
transition amplitude, and thus to $1/m^8$, we find
\be \Gamma \sim \frac{T^9}{m^8} \, . \label{decrate} \ee
In a typical process each $\phi$ carries off energy $\sim T$, and
thus the total energy loss per unit time of a star due to the
$\phi$ emission is $\dot E_T \sim - N \frac{T^{10}}{m^8} \sim -
\frac{M_S}{m_p} \frac{T^{10}}{m^8}$, where $N \sim M_S/m_p$ is the
number of particles in a star of mass $M_S$. Because stars are
typically predominantly made up of hydrogen, $m_p$ is the proton
mass. Rather than analyzing all the sources of data, we merely
quote the strongest bound which comes from the supernova SN1987a.
In order to agree with the observations, the total output of
$\phi$'s cannot exceed the luminosity of about $10^{53} erg/s \sim
10^{32} GeV^2$. Since $M_{S} \sim M_{\odot} \sim 1.6 \times
10^{57} GeV$ and $T \sim 30 MeV$, requiring $\dot E_T \le 10^{32}
GeV^2$ we find
\be m \ge 30 GeV \, , \label{snebound}\ee
which is weaker than (\ref{gravabound}). Hence because of the
bound (\ref{gravabound}) the supernova cooling is not
significantly affected by $\phi$ emission. We note that similar
bounds were also obtained from considering Goldstone boson
interactions in braneworlds \cite{golds}. Although these theories
are different, the bounds are similar because of the Goldstone
boson equivalence theorem.

We now turn to the cosmology of the model. Let us restrict to the
spatially flat FRW backgrounds for now. Starting with the usual
metrics for $g_{\mu\nu}$, the line element defining the
graviton-inflaton geometry is
\be ds^2 = -dt^2 + a^2 d\vec x^2 \, . \label{canonical} \ee
The translational symmetries require $\partial_k \phi = 0$, and
hence using (\ref{action}) we find that the metric in which the
Standard Model fields dwell is
\be d\bar s^2 = -(1+ \frac{ \dot \phi^2}{m^4}  ) dt^2 + a^2 d\vec
x^2 = - d\bar t^2 + a^2(\bar t) d\vec x^2 \, , \label{frwmetrics}
\ee
where $d\bar t = dt \sqrt{1+ \dot \phi^2/m^4}$. In this case the
field equations (\ref{fieldeq1})-(\ref{fieldeq3}) reduce to
\ba
&& 3  H^2 = \frac{1}{M^2_P} \bigl(\rho_\phi
+ \frac{1}{\sqrt{1-U^2}} \bar \rho_{SM}\bigr) \, , \nonumber \\
&& \frac{\ddot a}{a} = - \frac{1}{6M^2_P} \Bigl(\rho_\phi+3 p_\phi
+ \frac{\bar \rho_{SM}}{\sqrt{1-U^2}}
+ 3 \sqrt{1-U^2} \bar p_{SM} \Bigr) \, , \nonumber \\
&&\frac{d {\bar \rho}}{d\bar t}  + 3 \bar H (\bar \rho + \bar p) =
0 \, , ~~~~~~~ \bar p_{SM} = \bar w \bar \rho_{SM} \, ,  \nonumber \\
&&\ddot \phi + 3H \dot \phi - \frac{{\bar \rho}_{SM}}{m^4
(1-U^2)^{3/2}} \Bigl( \ddot \phi - 3 H \bar w (1-U^2) \dot \phi
\Bigr) = 0 \, , \label{frweqs} \ea
where we are still employing the obvious notations without and
with bars to distinguish the quantities built from metrics
(\ref{canonical}) and (\ref{frwmetrics}), and bearing in mind that
$U^2 = - \dot \phi^2/m^4$ and $\rho_\phi=p_\phi = \dot \phi^2$.
Here we are approximating the Standard Model influences with a
perfect fluid, obeying the equation of state $\bar p_{SM} = \bar w
\bar \rho_{SM}$ for some $\bar w$. Note that the two next-to-last
equations can be immediately integrated to yield $\bar \rho_{SM} =
\bar \rho_{SM}^{0} ~\Bigl(\frac{a_0}{a} \Bigr)^{3(1+\bar w)}$,
where $\bar \rho_{SM}^0$ is the initial value of the Standard
Model energy density when the description based on (\ref{frweqs})
became valid.

Although the equations (\ref{frweqs}) look quite formidable, it is
very simple to deduce their qualitative properties. In the regime
$m^4 \la \dot \phi^2 \la \Lambda^4 \sim M^4_P$, one finds the
following inequalities:
\ba \frac{{\bar \rho}_{SM}}{\sqrt{1-U^2}} &\simeq& \frac{m^2}{\dot
\phi} {\bar \rho}_{SM} \ll m^4 < \rho_{\phi} \, ,
\nonumber \\
{\sqrt{1-U^2}} {{\bar p}_{SM}} &\simeq& \frac{\dot \phi}{m^2}
{{\bar p}_{SM}} \le m^2 \dot \phi < p_\phi = \rho_{\phi} \, ,
\nonumber \\
\frac{{\bar \rho}_{SM}}{m^4(\sqrt{1-U^2})^{1/2}} &\simeq&
\frac{m^2}{\dot \phi} \frac{{\bar \rho}_{SM}}{m^4} \ll 1 \, ,
\nonumber \\
\frac{{\bar \rho}_{SM}}{m^4(\sqrt{1-U^2})^{3/2}} &\simeq&
\frac{m^6}{\dot \phi^3} \frac{{\bar \rho}_{SM}}{m^4} \ll  1 \, .
\label{ineqs} \ea
Because of these inequalities, all of the Standard Model
contributions in (\ref{frweqs}) are completely subleading to the
$\dot \phi$ sources in the regime $m^4 \la \dot \phi^2 \la
\Lambda^4 \sim M^4_P$. This in fact is exactly a tell-tale sign of
inflation: the matter contributions become irrelevant as the
inflationary dynamics sets in. Substituting these inequalities in
(\ref{frweqs}) we find the simple equations
\ba && 3 H^2 = \frac{1}{M^2_P} \frac{\dot \phi^2}{2} \, ,
\nonumber \\
&& \ddot \phi + 3 H \dot \phi = 0 \, , \label{prho} \ea i.e.
precisely the equations of a cosmology dominated by a stiff fluid
$p_\phi = \rho_\phi = \dot \phi^2/2$. Assuming that initially the
universe started with Planckian curvatures, as in chaotic
inflation \cite{chaotic}, the solution is
\be a = a_0 \Bigl( \frac{t}{t_P} \Bigr)^{1/3} \, ,
~~~~~~~~~~~~~~~~ \phi = \phi_0 + \sqrt{\frac23} M_P
\ln\bigl(\frac{t}{t_P}\bigr) \, . \label{solns} \ee
Even though the geometry of (\ref{solns}) is the same as the
holographic cosmology background of \cite{tomwilly}, the
difference is that here the geometry is sourced by a simple scalar
field $\phi$ whereas in the context of holographic cosmology it
emerges in response to a black hole gas. Thus the fluctuations
around the background would be very different. Now, because $\dot
\phi = \sqrt{\frac23} M_P /t$, the matter frame metric is, using
(\ref{frwmetrics}),
\be d \bar s^2 = - (1+ \frac23 \frac{M^2_P}{m^4 t^2}) dt^2 + a_0^2
\Bigl(\frac{t}{t_P}\Bigr)^{2/3} d\vec x^2 \, , \label{matrsoln}
\ee
or therefore, using $d\bar t = dt \sqrt{1+ \frac23
\frac{M^2_P}{m^4 t^2}} \simeq \sqrt{\frac23} \frac{M_P}{m^2}
~dt/t$,
\be d\bar s^2 = -d\bar t^2 + a^2_0 e^{\frac23\frac{m^2}{ M_P} \bar
t} d \vec x^2 \, , \label{infl}\ee
i.e. precisely the slow-roll inflation, with an almost constant
Hubble parameter,
\be \bar H \simeq \frac{m^2}{\sqrt{6} M_P} \, . \label{hubble} \ee
Thus the effective field theory description of $\phi$-dominated
cosmology is a low scale inflation. An indication of the presence
of an inflationary attractor in theories which in the matter frame
metric contain similar operators to those present here was noted
in \cite{amendola}. When $m \sim m_{EW} \sim TeV$, the Hubble
scale during inflation is $\bar H \sim {\rm mm}^{-1}$, i.e.
$V_{eff} \sim TeV^4$, corresponding to $TeV$ scale inflation as in
the examples of \cite{lowscale}. We stress that this mechanism of
inflation is different than the so-called {\tt k}-inflation
\cite{kinfl}. This can be readily seen by rewriting the theory
(\ref{action}) completely in terms of the metric $\bar g_{\mu\nu}$
to which the matter couples, and noting that it contains operators
$\propto \frac{M^2_P}{m^4} \partial_\mu \phi
\partial_\nu \phi {\bar R}^{\mu\nu}$, which play a key role here
and are absent in {\tt k}-inflation.

The inflationary stage terminates gracefully because as the time
goes on, $\dot \phi \sim M_P/t$ decreases. When it reaches $m^2$,
inflation ceases. Indeed, in the regime $\dot \phi^2 < m^4$,
because $\phi$ is completely without a potential, its energy
density scales as $1/a^6$, and so the scale factor rapidly changes
behavior, scaling as some low power of $t$ after inflation, while
$\dot \phi$ is diluted very fast. It is straightforward to
determine the duration of the inflationary phase. Using the form
of the solution (\ref{solns}) during the inflationary regime, we
can rewrite the scale factor as a function of $\dot \phi$: $a =
a_0 (\frac{\dot \phi_0}{\dot \phi})^{1/3}$, where $\dot \phi_0
\sim M^2_P$ is the initial value of the inflaton gradient. Thus,
the total amount of inflation is given by the number of e-folds
\be {\cal N} = \ln \frac{a_{exit}}{a_0} \simeq \frac13 \ln
\frac{\dot \phi_0}{m^2} \, . \label{efolds} \ee
Taking the initial condition for the inflaton to correspond to the
Planckian energy density, $\rho_0 \simeq \dot \phi^2_0 \sim M^4_P$
\cite{chaotic}, and choosing $m \ga m_{EW} \sim TeV$ to saturate
(\ref{gravabound}) we get
\be {\cal N} \simeq \frac23 \ln \frac{M_P}{m} \la 25 \, .
\label{nomefolds} \ee
This suffices to solve the horizon problem if the reheating
temperature after inflation is $\sim MeV$, because inflation
started late, with the initial horizon size $\sim \bar H^{-1} \sim
{\rm mm}$. In this case  the formula linking the number of e-folds
needed for the post-inflationary entropy production to the
reheating temperature and the scale of inflation
\cite{kolbturner}, ${\cal N} \simeq 67 - \ln(M_P/m) - \frac13 \ln
(m/T_{RH}) \, ,$ gives exactly ${\cal N} \simeq 25$ for $m \sim
m_{EW} \sim TeV$ and $T_{RH} \sim MeV$, agreeing with
(\ref{nomefolds}). If $m > m_{EW}$, inflation would be shorter,
reducing its efficiency for solving the horizon problem.

The processes of reheating and generation of the primordial
density fluctuations are somewhat involved. We first discuss the
nature of the problems, and then turn to a specific solution based
on another light field \cite{curv}. Since the inflaton energy
density after inflation scales as $1/a^6$, after inflation the
cosmological evolution rapidly falls under the control of the
matter sector. However, inflaton reheating is very inefficient.
The symmetries $\phi \rightarrow \phi + {\cal C}$, $\phi
\leftrightarrow - \phi$ which protect the inflaton from the
Standard Model corrections prevent strong inflaton-matter
couplings and hamper reheating. This is similar to other
non-oscillatory models of inflation, where the inflaton after
inflation does not fall into a minimum of a potential
\cite{nomod}. In fact, this model is an extreme non-oscillatory
model, because the matter couples only to the metric ${\bar
g}_{\mu\nu}$. Thus the only particle production is gravitational,
driven by the evolution of the vacuum of the quantum field theory
of matter in the ${\bar g}_{\mu\nu}$ background. This means that
the reheating temperature is given by \cite{ford}
\be T_{RH} \sim \bar H \sim \frac{m^2}{M_P} \, . \label{gravreh}
\ee
Requiring $T_{RH} \ga MeV$ in order to have nucleosynthesis, one
needs $m \ga 10^5 TeV$, reducing the number of e-folds to $\sim
17$. This is too few e-folds to accommodate a solution of the
horizon problem, but could be useful for other model building
purposes.

In the usual potential-driven models of inflation, the inflaton
quantum fluctuations on the potential plateau generate density
perturbations, which later serve as the seeds for structure
formation in the post-inflationary universe \cite{bard,perts}. The
key reason why this mechanism for generation of density
perturbations is successful is that during inflation the inflaton
fluctuations are imprinted on the background as curvature
inhomogeneities which are stretched to scales greater than the
apparent horizon, where they freeze out: their amplitude rapidly
approaches a constant value, leading to $\delta \rho/\rho \sim
H^2/\dot \phi \sim {\rm const.}$ Thus the resulting spectrum is
scale-invariant, fixed by inflationary dynamics, and protected
from the details of subsequent evolution.

Our analysis shows that this does not happen with the $\phi$-field
fluctuations in this model, contrary to the claims of \cite{vsl}.
The problem is that in this case the curvature perturbations never
freeze out. The reason is quite simple: the canonically normalized
fluctuations couple to the background metric $g_{\mu\nu}$, and the
background is given by (\ref{solns}), with $a \sim t^{1/3}$. This
is a decelerating geometry, singular at $t=0$ and with the
apparent horizon given by $l_H = 3 t$, which is spacelike. Since
the wavelength of the fluctuations obeys $\lambda = \lambda_0
a/a_0 = \lambda_0 (t/t_0)^{1/3}$, it grows more slowly than the
horizon $l_H$ as time goes on. The evolution of the horizons and a
characteristic wavelength is given in Fig. 3).
\begin{figure}[thb!]
 \hspace{3.2truecm}
\vspace{-1.5truecm} \epsfysize=3.8truein \epsfbox{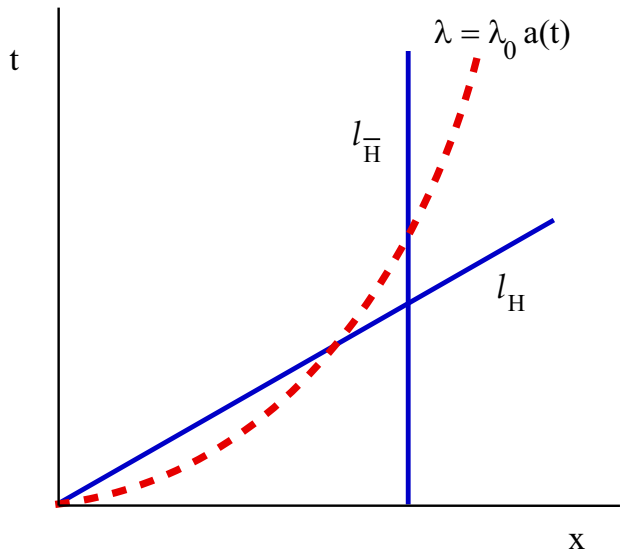}
\caption[]%
{\small\sl Apparent horizons ${l}_{H}=H^{-1}$, ${l}_{\bar H} =
{\bar H}^{-1}$ and the wavelength $\lambda$ of a typical
fluctuation of $\phi$ as functions of time $t$. }
\end{figure}
So a fluctuation which originates inside the horizon $l_H$ remains
inside of it forever. The fact that its wavelength will become
greater than the apparent horizon in the matter frame, $l_{\bar H}
= \bar H^{-1}$, is of little dynamical consequence since in this
frame the fluctuations do not have canonical kinetic terms. This
only serves to set the proper normalization for the momenta, and
wavelengths, of the fluctuations after inflation. The
perturbations which are generated from the quantum fluctuations of
$\phi$ will therefore continue being redshifted away, and will end
up being exponentially small.

To see this we use the gauge-invariant cosmological perturbation
theory \cite{bard,perts}. Although the dynamics of the
perturbations will be governed by a linearized theory which will
differ significantly from the standard perturbation theory during
inflation, because of the couplings in (\ref{metrics}), because
the theory is generally covariant we can use the formalism of
\cite{bard} to identify the gauge invariant potentials. In the
longitudinal gauge the background+perturbations are defined by
\ba ds^2 &=& a^2(\eta) \Bigl[ -\Bigl(1+2{ \Psi }(\eta,\vec
x)\Bigr) d\eta^2 + \Bigl(1+2{\Phi}(\eta,\vec x)\Bigr) d\vec x^2
\Bigr] \, , \nonumber \\
\phi &=& \phi(\eta) + \delta \phi(\eta, \vec x) \, .
\label{flucfrw} \ea
The conformal time $\eta$ is related to the usual comoving FRW
time $t$ by $dt = a d\eta$, which gives $t = t_P (\frac{3\eta}{2
t_P})^{3/2}$, and $a(\eta), \phi(\eta)$ are obtained from
(\ref{solns}). The potentials $\Phi, \Psi$ and the inflaton
perturbation $\delta \phi$ are related by momentum conservation as
$\Psi = - \Phi$, $\phi' \delta \phi = - 2 m^2_P (\Phi' + {\cal
H}\Phi)$, where ${\cal H} = a'/a$, and primes denote derivatives
with respect to $\eta$. One then defines the curvature
perturbations as the perturbations on the isodensity spatial
slices. In terms of the gauge-invariant potential $\Theta = \Phi -
\frac{\cal H}{\phi'} \delta \phi $ they are
\be \frac{\delta {\cal R}_3}{R} = \frac{1}{3a^2 H^2} \vec \nabla^2
\Theta(\eta,\vec x) \, .
\label{totpert} \ee
The canonically normalized scalar field corresponding to this
perturbation is
\be \varphi = a \delta \phi - \frac{a \phi'} {\cal H} \, \Phi = -
{\cal Z} \Theta  \, .\label{scalarinfl} \ee
Following a common practice we have defined $ {\cal Z} = \frac{a
\phi'} {\cal H}= \frac{a \dot \phi} {H}$ \cite{perts}. Expanding
in Fourier modes, and using the definition of the power spectrum
${\cal P}(k)\delta^{(3)}(\vec k - \vec q) = \frac{k^3}{2\pi^2}
\langle \Theta_{\vec k}(\eta) \, \Theta^\dagger{}_{\vec q}(\eta)
\rangle $ yields
\be
{\cal P}(k)\delta^{3}(\vec k - \vec q) = \frac{k^3}{2\pi^2}
\Bigl(\frac{H}{\dot \phi}\Bigr)^2 \langle \frac{\varphi_{\vec k}}{a} \,
\frac{\varphi^\dagger{}_{\vec q}}{a} \rangle \, ,
\label{power}
\ee
where $\langle {\cal O} \rangle$ stands for the quantum
expectation value of the 2-point operator ${\cal O}$ in the
quantum state of inflation. The curvature perturbation in the
gravitational frame is $\Bigl(\frac{\delta \rho}{\rho} \Bigr) \sim
{\cal P}^{1/2}(k)$.

However, we are interested in the curvature perturbation as seen
in the matter frame, in terms of the variables adopted to the
metric $\bar g_{\mu\nu}$. From the relation (\ref{metrics}) and
the background solution (\ref{solns}) it is straightforward albeit
tedious to compute the relation of the curvature perturbations in
the gravitational frame $\Theta$ and the matter frame $\bar
\Theta$. Keeping terms up to linear order in perturbations,
redefining the conformal time according to $d \bar \eta =
\sqrt{1+\phi'^2/m^4} d\eta$, and then performing an infinitesimal
diffeomorphism $d\bar \eta \rightarrow d\bar \eta + \frac{\phi'
\delta \phi}{a^2 m^4 \sqrt{1+\phi'^2/m^4}}$, we find that in the
limit $\dot \phi^2 \gg m^4$, valid during inflation, they obey the
relationship
\be \bar \Theta = \Theta +  \frac{m^4}{\dot \phi^2} \frac{\cal
H}{\phi'} \delta \phi + \frac{1}{3 a m^2} \frac{d }{d\bar
\eta}(\frac{m^4}{\dot \phi^2} \delta \phi ) + {\cal
O}\Bigl((\frac{m^4}{\dot \phi^2})^2 \Bigr) \, .\label{flucfrwm}
\ee
Hence to the leading order $\bar \Theta = \Theta$, and so we can
simply compute $\Theta$ in the gravitational frame, where the
scalar and graviton modes have canonical kinetic terms, and carry
over the result to the matter frame. The main difference between
the frames arises because in the matter frame we should compare
the curvature perturbation $\Theta$ to the background curvature of
the matter frame metric $\bar g_{\mu\nu}$. This yields
\be  \frac{\delta \bar \rho}{\bar \rho}  \simeq
\Bigl(\frac{R}{\bar R} \Bigr)^{1/2}  \frac{\delta \rho}{\rho}
\simeq \Bigl(\frac{H}{\bar H} \Bigr)  \frac{\delta \rho}{\rho} \,
. \label{pertsmat} \ee
We can now estimate the perturbations in
the long wavelength limit.

Since (\ref{solns}) implies that ${\cal Z}''/{\cal Z} = -
\frac{1}{4\eta^2}$ to the leading order, one sees that the Fourier
modes of $\varphi$ obey the field equation
\be \varphi_{\vec k}'' + \bigl(k^2 + \frac{1}{4 \eta^2} \bigr) \,
\varphi_{\vec k} = 0 \, , \label{scalfe} \ee
with the solutions
\be {\varphi}_{\vec k} = \sqrt{\frac{\eta}{t_P}} \Bigl( A_{\vec k}
J_0(k \eta) + B_{\vec k} Y_0(k \eta) \Bigr) \, ,
\label{eigensolns} \ee
where $J_0, Y_0$ are Bessel functions of index zero. Because
inflation progresses as $t$ grows, the long wavelength limit
behavior of the modes is encoded in the limit $k \eta \gg 1$.
Because in this limit
\be J_0 \rightarrow \sqrt{\frac{2}{\pi k \eta}} \cos(k \eta -
\frac{\pi}{4}) \, , ~~~~~~~~~~~~~ Y_0 \rightarrow
\sqrt{\frac{2}{\pi k \eta}} \sin(k \eta - \frac{\pi}{4}) \, ,
\label{bessels} \ee
the mode functions behave as
\be {\varphi}_{\vec k} \rightarrow \frac{1}{\sqrt{k}} \Bigl(
a_{\vec k} e^{-ik \eta} + a^\dagger_{\vec k} e^{ik \eta} \Bigr) \,
, \label{eigensols} \ee
where we have defined $a_{\vec k}, a^\dagger_{\vec k}$ from
$A_{\vec k}, B_{\vec k}$ in an obvious way. Therefore in the limit
$k \eta \gg 1$ after a simple algebra we get
\be {\cal P}(k) \rightarrow \frac{k^2}{M^2_P} n(k)
\Bigl(\frac{t_P}{t}\Bigr)^{2/3} \, , \label{powerres} \ee
and therefore, using (\ref{solns}) and (\ref{pertsmat}) and
defining the physical momentum of the fluctuations $p = k/a$,
\be \frac{\delta \bar \rho}{\bar \rho} \simeq \frac{p}{\bar H}
n^{1/2}(p) e^{-3 \bar H \bar t} \, . \label{matpert} \ee
Here $n(p)$ is the occupation number of modes as a function of
their momentum in the initial state of inflation $\langle ~
\rangle$. From this formula we see that unless the initial state
of $\phi$ is very precisely fine-tuned such that $n(p) =
\alpha/p^2$, the spectrum of fluctuations will not be flat. This
is very hard to justify because in the limit $t \rightarrow t_P$
the solution is singular, and the fluctuations of $\phi$ are
random, and very large. More importantly, the curvature
fluctuations do not freeze out as inflation proceeds. Indeed, if
inflation lasts ${\cal N} = \bar H \bar t_{exit} \sim 25$ e-folds,
the amplitude of density perturbations in horizon-size modes,
which were the first to leave the matter-frame apparent horizon
$l_{\bar H}$, is diluted by the factor $(e^{25})^3 = e^{75} \sim
10^{32}$ by the time inflation terminates: $\delta \bar \rho/\bar
\rho \sim 10^{-32}$. Thus these fluctuations are much too small
when compared to the COBE amplitude $\delta \bar \rho/\bar \rho
\sim 10^{-5}$, and cannot give rise to the observed structure in
the universe. However, at least they do not destabilize inflation
once it sets in.

A cure to the problems of reheating and generation of density
perturbations may be the curvaton mechanism \cite{curv}. In this
case the curvaton should be a very light field $\sigma$, with a
mass $\mu \la \bar H \sim m^2/M_P$. If it is stuck at a large {\it
vev} initially, say $\sigma \sim M_P$, it will remain there all
the way through inflation. It will give rise to a small
cosmological term, $\sim \mu^2 M^2_P \la m^4$, which however does
not significantly affect the background as we have discussed
following the equation (\ref{exppowers}). Once inflation
terminates and $\bar H$ starts to decrease, $\sigma$ will begin to
roll towards its minimum. It will have initial energy density
$\rho_\chi \sim \mu^2 \sigma^2_0 \sim \bar H^2 M^2_P \sim m^4$,
and will immediately take over the control of the cosmological
evolution from $\phi$, scaling like cold dark matter. It could
reheat the post-inflationary universe efficiently if it couples to
a fermion field $\psi$ in the matter sector with a Yukawa coupling
\be (m_{\psi} - g \sigma) \bar \psi \psi \, . \label{yukawa} \ee
As the field $\sigma$ moves towards the minimum, within a time
$\sim 1/\mu$ it will scan all possible values. When it reaches
$\chi = m_{\psi}/g$ it will copiously preheat the fermions $\psi$
through the parametric resonance phenomenon \cite{lyova}. One can
give a crude estimate of the number density of fermions $n_\psi$
produced in this way by recalling that the decay rate of $\chi$
into two fermions is $\Gamma \sim g^2 \mu$, and that the number
density of $\chi$s is $n_\chi \simeq \mu \sigma^2$, so that using
the continuity equation for the fermion number density,
$\frac{1}{a^4} \frac{d}{d\bar t} (a^4 n_\psi) \sim \Gamma n_\chi$
\cite{lyova}. The fermion production lasts a fraction of $1/\mu$
so that the resulting fermion number density is
\be n_\psi \simeq \mu m^2_\psi \, . \label{fernumb} \ee
When the field $\sigma$ settles down in the minimum, the fermion
energy density will be $\rho_\psi \sim n_\psi m_\psi \sim \mu
m^3_{\psi}$. The fermions $\psi$ need to quickly decay into the
Standard Model particles to complete the reheating process. Taking
$m_\psi \sim m$, the reheating temperature is given by
\be T_{RH} \sim \Bigl( \mu m^3_\psi \Bigr)^{1/4} \sim
\Bigl(\frac{m}{M_P} \Bigr)^{1/4} m \, , \label{reheatt} \ee
or $T_{RH} \la 100 MeV$ if $m \sim m_{EW} \sim TeV$, which may be
sufficient to have conditions for a successful nucleosynthesis.
The proper treatment of nonlinear effects may further enhance the
reheating efficiency \cite{lyova}. There may also be other
possibilities for curvaton reheating, as discussed in
\cite{curvreh}.

The same field may also produce the required density fluctuations.
During inflation, because the curvaton dwells in the matter frame
geometry defined by $\bar g_{\mu\nu}$, its fluctuations freeze out
just like the fluctuations of any light scalar during inflation.
They obey a field equation \cite{curv}
\be \frac{d^2 \sigma_{\vec k}}{d\bar t^2} + 3 {\bar H}
\frac{d\sigma_{\vec k}}{d\bar t} + (\frac{k^2}{a^2(\bar t)} +
\mu^2) \sigma_{\vec k} = 0 \, , \label{curvfluc} \ee
and thus in the limit $k^2/a^2 \ll 1$ they yield $\sigma_{\vec k}
\rightarrow \alpha_{\vec k} + \beta_{\vec k}/a^3$. The
perturbations are Gaussian, and start off as isocurvature
perturbations, which however are converted into adiabatic
perturbations after inflation \cite{curv}, giving a nearly
scale-invariant spectrum with the amplitude
\be {\cal P}^{1/2} \sim r \frac{\bar H}{\pi \sigma_*} \, ,
\label{curvperts} \ee
where $r$ is the curvaton fraction of the total energy density
after inflation, and $\bar H$ and $\sigma_*$ are the values of the
Hubble parameter and the curvaton near the end of inflation. In
the case of low scale inflation, one must also ensure that the
curvaton mass changes rapidly after inflation in order not to
spoil nucleosynthesis \cite{lyth}, but such models are possible in
principle.

So far we have been ignoring the curvature problem. The model of
inflation discussed here does not solve it. This is reminiscent of
other models of low scale inflation. None of them are stand-alone
solutions of the curvature problem. If inflation begins at a scale
$\bar H \ll M_P$, something else must have kept the universe from
collapsing until it reached the age $\sim \bar H^{-1}$, where late
inflation can begin. Thus for low scale inflation to start, ${\tt
k}/a^2$ must be very small. However, since the solution
(\ref{solns}) is decelerating when expressed in terms of the
metric $g_{\mu\nu}$, the curvature problem here is more severe
than in other low scale inflation models. To see this, note that
in order to get a sufficiently small curvature today, we must
ensure roughly ${\tt k}/a^2_{now} \la H^2_0/100$, and therefore at
the end of inflation the curvature term must satisfy ${\tt
k}/a^2_{exit} \la m H_0/100$. Hence using the solution
(\ref{solns}), we see that since initially $H \la M_P$, we must
have ${\tt k}/(a_0^2 M_P^2) \la (M_P/m)^{4/3} (m H_0/M^2_P)/100
\sim 10^{-57}$, which is a bit better than the required amount of
fine tuning without any inflation. However one still needs to
explain the origin of such a small number. Problems with curvature
were noticed in \cite{visser}. There are several different
possibilities for ensuring that ${\tt k}/a^2$ is small at the
onset of low scale inflation.  One possibility is to have an early
stage of inflation, driven by some other gravitationally coupled
scalar, followed by the low scale inflation \cite{ahdkmr}. Such
models might arise in the context of Little String Theories at a
$TeV$ \cite{adg}, where the early inflation would take a
Planck-scale universe and blow it up to $TeV^{-1}$ size while
generating the Planck-electroweak hierarchy. Another possibility
might be the holographic cosmology \cite{tomwilly}. In that case,
the very early universe would start in the most entropically dense
state, with a nearly vanishing curvature, which would evolve
towards the regime of low density where conventional evolution can
take place \cite{tomwilly}.

In closing, we have shown that a theory of gravity based on a
special case of Bekenstein's disformal couplings, where matter
couples to a combination $g_{\mu\nu} - \frac{1}{m^4} \partial_\mu
\phi \partial_\nu \phi$, gives rise to an epoch of short, low
scale inflation. Here $\phi$ is a pseudoscalar field, which is
invariant under $\phi \rightarrow \phi + {\cal C}$, $\phi
\leftrightarrow - \phi$. These symmetries protect the inflaton
sector from the Standard Model radiative corrections. The leading
order coupling of $\phi$ to the matter fields is via the universal
dimension-8 operator $\frac{1}{m^4} \partial_\mu \phi
\partial_\nu \phi \, T^{\mu\nu}$. Because of such couplings,
the presence of $\phi$ is not in violation of any experimental
bounds, even though it is massless, and may even lead to new
signatures accessible to future collider experiments when $m \sim
TeV$. For that value of the mass $m$ the number of e-folds of
inflation is ${\cal N} \sim 25$, which is just enough to solve the
horizon problem. The fluctuations of $\phi$ do not give rise to
the satisfactory spectrum of density perturbations, and its
couplings are too weak for efficient reheating, but both of these
problems can be solved by adding other light scalar(s) such as the
curvaton \cite{curv}. It would be interesting to explore further
implications of this mechanism and see if it can arise from some
fundamental theory.

\vspace{1.5cm} {\bf Acknowledgements}

We thank A. Albrecht, N. Arkani-Hamed, T. Banks, S. Dimopoulos, J.
Erlich, D.~E. Kaplan, M. Kaplinghat, E. Katz, A. Linde, A. Nelson,
Y.-S. Song, L. Sorbo and L. Yaffe for useful discussions. This
work was supported in part by the NSF Grant PHY-0332258 and in
part by a Research Innovation Award from the Research Corporation.
The author thanks the KITP at UCSB for hospitality during this
work and for partial support by the NSF Grant PHY-99-07949, and
the Harvard Theory Group for hospitality during the completion of
this work.

\end{document}